\documentclass[conference]{IEEEtran}
\usepackage[utf8]{inputenc}
\usepackage[T1]{fontenc}

\usepackage[usenames,dvipsnames]{xcolor}
\usepackage{amssymb,amsmath,amsfonts}
\usepackage{tikz,pgffor}
\usetikzlibrary{arrows}
\usetikzlibrary{shapes}
\usetikzlibrary{calc}
\usetikzlibrary{automata}
\usepackage{txfonts}
\usepackage{wrapfig}
\usepackage{cutwin}
\usepackage{stackrel}
\usepackage{enumitem}
\usepackage{eurosym}
\usepackage{multirow}

\usepackage[linesnumbered,ruled,noend]{algorithm2e}

\usepackage{microtype}

\tikzstyle{ran}=[circle,thick,draw,minimum size=1.4em,inner sep=0em]
\tikzstyle{act}=[rectangle,thick,draw,minimum size=1.4em,inner sep=1ex]
\tikzstyle{tran}=[thick,draw,->,>=stealth]

\newcommand{\Nset}{\mathbb{N}}
\newcommand{\Nseto}{\Nset_0}

\newcommand{\Rsetp}{\mathbb{R}_{>0}}
\newcommand{\Rsetpo}{\mathbb{R}_{\ge 0}}

\renewcommand{\vec}[1]{\mathbf{#1}}

\newcommand{\eps}{\varepsilon}
\newcommand{\dist}{\mathcal{D}}

\newcommand{\probm}{\mathrm{Pr}}

\newcommand{\expred}{E}

\newcommand{\strategy}{policy}
\newcommand{\strategies}{policies}
\newcommand{\strategyEvaluation}{policy evaluation}
\newcommand{\strategyImprovement}{policy improvement}
\newcommand{\valueVector}{\mathbf{x}}

\newcommand{\roots}{R}
\newcommand{\candidates}{L}
\newcommand{\func}{f_{\sta, \valueVector}}
\newcommand{\pol}{p_{\sta, \valueVector}}
\newcommand{\poll}{q_{\sta, \valueVector}}
\newcommand{\delayVector}{delay function}

\renewcommand{\psi}{\text{Pois}}

\newcommand{\code}[1]{\texttt{#1}}

\newcommand{\fdC}{C}

\newcommand{\states}{S}
\newcommand{\sta}{s}

\newcommand{\goalStates}{G}

\newcommand{\initstate}{\sta_{in}}

\newcommand{\gen}{Q}
\newcommand{\prob}{\mathrm{P}}

\newcommand{\allact}{\states_{\!\mathrm{fd}}}

\newcommand{\paramvalspace}{D}

\newcommand{\trans}{\mathrm{F}}

\newcommand{\delays}{d}
\newcommand{\timeouts}{\mathbf{d}}

\newcommand{\nodel}{\infty}

\newcommand{\rateRew}{\mathcal{R}}
\newcommand{\impRew}{\mathcal{I}}
\newcommand{\impRewExp}{\mathcal{I}_\prob}
\newcommand{\impRewFix}{\mathcal{I}_\trans}
\newcommand{\costFdC}{\mathit{Cost}}

\newcommand{\expectedRewExp}{\mathcal{J}_\gen}
\newcommand{\expectedRewFix}{\mathcal{J}_\trans}

\DeclareMathOperator*{\argmin}{argmin}

\newcommand{\Sset}{\states_{\!\mathrm{set}}}
\newcommand{\Soff}{\states_{\!\mathrm{off}}}
\newcommand{\Tval}{\mathit{Dval}}
\newcommand{\ac}[1]{\langle #1 \rangle}

\newcommand{\vtx}{\sta}

\newcommand{\mtran}{T}

\newcommand{\mcost}{\text{\euro}}

\newcommand{\mdp}{\mathcal{M}}

\newcommand{\ctrun}{\omega}

\newcommand{\sleep}{\mathit{sleep}}
\newcommand{\busy}{\mathit{busy}}
\newcommand{\idle}{\mathit{idle}}

\ifCLASSINFOpdf
\else
\fi

\newtheorem{theorem}{Theorem}

\newtheorem{example}[theorem]{Example}
\newtheorem{proposition}[theorem]{Proposition}

\newcommand{\theoremlike}[2]{\par\medskip\penalty-250\refstepcounter{theorem}{{\itshape#2
\ref{#1}:}}}

\newcommand{\problemlike}[2]{\par\medskip\penalty-250\refstepcounter{theorem}{{\itshape#2
\ref{#1}:}}}

\newcommand{\probhelperpre}[2]{\problemlike{#1}{#2}}
\newcommand{\thmhelperpre}[2]{\theoremlike{#1}{#2}}
\newcommand{\thmhelperpost}{\par\medskip}

\begin{document}
\title{Efficient Timeout Synthesis in Fixed-Delay CTMC Using Policy Iteration}

\author{\IEEEauthorblockN{L\kern-.8ex'ubo\v{s} Koren\v{c}iak\IEEEauthorrefmark{1},
Anton\'{\i}n Ku\v{c}era\IEEEauthorrefmark{1},
Vojt\v{e}ch \v{R}eh\'ak\IEEEauthorrefmark{1}}
\IEEEauthorblockA{\IEEEauthorrefmark{1}Faculty of Informatics, Masaryk University, Brno, Czech Republic\\Email: \{korenciak, kucera, rehak\}@fi.muni.cz}}

\maketitle

\begin{abstract}
We consider the fixed-delay synthesis problem  for continuous-time Markov chains extended with fixed-delay transitions (fdCTMC). The goal is to synthesize concrete values of the fixed-delays (timeouts) that 
minimize the expected total cost incurred before reaching a given set of target states. The same problem has been considered and solved in previous works by computing an optimal policy in a certain discrete-time Markov decision process (MDP) with a huge number of actions that correspond to suitably discretized values of the timeouts. 

In this paper, we design a \emph{symbolic} fixed-delay synthesis algorithm which avoids the explicit construction of large action spaces. Instead, the algorithm computes a small sets of ``promising'' candidate actions on demand. The candidate actions are selected by minimizing a certain objective function by computing its symbolic derivative and extracting a univariate polynomial whose roots are precisely the points where the derivative takes zero value. Since roots of high degree univariate polynomials can be isolated very efficiently using modern mathematical software, we achieve not only drastic memory savings but also speedup by three orders of magnitude compared to the previous methods.
\end{abstract}

\section{Introduction}
\label{sec-intro}

Continuous-time Markov chains (CTMC) are a fundamental formalism widely used in performance and dependability analysis. CTMC can model exponentially distributed events, but not \emph{fixed-delay} events that occur after a fixed amount of time with probability one\footnote{A fixed-delay distribution is a  typical example of a distribution where the standard phase-type approximation technique~\cite{Neuts:book} produces a large error unless the number of auxiliary states is very large; see, e.g., \cite{KKR:EPEW2014,fackrell2005fitting}.}. Since fixed-delay events are indispensable when modeling systems with \emph{timeouts} (i.e., communication protocols~\cite{TTC-book}, time-driven real-time schedulers~\cite{scheduling-in-real-time}, etc.), a lot of research effort has been devoted to developing formalisms that generalize CTMC with fixed-delay transitions. Examples include deterministic and stochastic Petri nets~\cite{marsan1987petri}, delayed CTMC~\cite{guet2012delayed}, or fixed-delay CTMC (fdCTMC) \cite{KKR:EPEW2014,BKKNR:QEST2015,KRF:iFM2016}. 

In practice, the duration of fixed-delay events (timeouts) is usually determined ad-hoc, which requires a considerable amount of effort and expertise. Hence, a natural question is whether the (sub)optimal timeouts can be synthesized \emph{algorithmically}. For fdCTMC, an algorithm synthesizing suboptimal timeouts was given in  \cite{BKKNR:QEST2015}. This algorithm is based on explicitly constructing and solving a discrete-time Markov decision process (MDP) whose actions correspond to suitably discretized admissible timeout values. Since the number of these actions is always large, the applicability of this algorithm is limited only to small instances for fundamental reasons. 
\smallskip

\noindent
\textbf{Our contribution.} In this paper, we design a new \emph{symbolic} algorithm for synthesizing  suboptimal timeouts in fdCTMC up to an arbitrary small error. Although we build on the results of \cite{BKKNR:QEST2015}, the functionality of our algorithm is different. First, the explicit construction of the aforementioned discrete-time Markov decision process is completely avoided, which drastically reduces memory requirements. Second, the search space of the underlying policy improvement procedure is restricted to a small subset of ``promising candidates'' obtained by identifying the local minima of certain analytical functions constructed ``on-the-fly''. This allows to safely ignore most of the discretized timeout values, and leads to speedup by three orders of magnitude compared to the algorithm of~\cite{BKKNR:QEST2015}. Consequently, our algorithm can synthesize suboptimal timeouts for non-trivial models of large size (with more than 20000 states) which would be hard to obtain manually.
\smallskip

Now we explain our results and their relationship to the previous works in greater detail. This requires a more precise understanding of the notions mentioned earlier, and therefore we switch to a semi-formal level of presentation. The rest of this introduction is structured as follows:
\begin{itemize}
	\item In Section~\ref{sec-fdCTMC}, we introduce the fdCTMC formalism, explain its semantics, and formalize the objective of fixed-delay synthesis.
	\item In Section~\ref{sec-our-alg}, we describe the key ingredients of our algorithm in more detail.
	\item In Section~\ref{sec-related}, we explain the relationship to previous works.  
\end{itemize}
In Sections~\ref{sec-fdCTMC} and~\ref{sec-our-alg}, some technical details unnecessary for basic understanding of the presented results are omitted. These can be found in Section~\ref{sec-algorithm}, where we assume familiarity with the notions introduced in Sections~\ref{sec-fdCTMC} and~\ref{sec-our-alg}. The experimental outcomes are presented in Section~\ref{sec:experiments}.

\subsection{Fixed-delay CTMC and the objective of fixed-delay synthesis}
\label{sec-fdCTMC}
A fdCTMC is a tuple $(S,\lambda,\prob,\allact,\trans)$, where $S$ is a finite set of states, $\lambda \in \Rsetpo$ is a common exit rate for the states, $\prob\in \Rsetpo^{S\times S}$ is a stochastic matrix specifying the probabilities of ``ordinary'' exp-delay transitions between the states, $\allact\subseteq S$ is a subset of states where a fixed-delay transition is enabled, and $\trans\in \Rsetpo^{S\times S}$ is a stochastic matrix
such that $\trans(s,s) = 1$ for all $s \in S \smallsetminus \allact$. For the states of $\allact$, the matrix $\trans$ specifies the probabilities of fixed-delay transitions. The states of $S \smallsetminus \allact$ are declared as absorbing by $\trans$, which becomes convenient in Section~\ref{sec-alg-full}. In addition, we specify a \emph{delay function} $\timeouts:\allact\rightarrow \Rsetp$ which assigns a concrete delay (timeout) to each state of~$\allact$. Note that $(S,\lambda,\prob)$ is an ``ordinary'' CTMC where the time spent in the states of $S$ is determined by the exponential distribution with the same\footnote{We can assume without restrictions that the parameter $\lambda$ is the same for all states of $S$, because every CTMC can be effectively transformed into an equivalent CTMC satisfying this property by the standard uniformization method; see, e.g., \cite{Norris:book}. Note that the transformation causes zero error.} parameter~$\lambda$.

The fdCTMC semantics can be intuitively described as follows. Imagine that the underlying CTMC $(S,\lambda,\prob)$ is now equipped with an alarm clock. When the alarm clock is turned off, our fdCTMC behaves exactly as the underlying CTMC. Whenever a state $s$ of $\allact$ is visited and the alarm clock is off at the time, it is turned on and set to ring after $\timeouts(s)$ time units. Subsequently, the process keeps behaving as the underlying CTMC until either a state of $S\smallsetminus \allact$ is visited (in which case the alarm clock is turned off), or the accumulated time from the moment of turning the alarm clock on reaches the value when the alarm clock \emph{rings} in some state $s'$ of $\allact$. In the latter case, an outgoing fixed-delay transition of $s'$ takes place, which means that the process changes the state randomly according to the distribution $\trans(s',\cdot)$, and the alarm clock is either newly set or turned off (depending on whether a state of $\allact$ or $S\smallsetminus \allact$ is entered, respectively).
\smallskip

\begin{figure}[t]\centering
 	\begin{tikzpicture}[x=1.5cm,y=1.5cm,font=\scriptsize]
 	\node (A) at (0,0)   [ran] {$A$};
 	\node (B) at (1,0)   [ran] {$B$};
 	\node (C) at (2,0)   [ran] {$C$};
 	\node (F) at (1,-1)  [ran] {$F$};
 	\draw [tran,->] (A) -- node[above] {$p$} (B);
 	\draw [tran,->] (A) -- node[right, near start] {$1{-}p$} (F);
 	\draw [tran,->] (B) -- node[above] {$p$} (C);
 	\draw [tran,->] (B) -- node[right] {$1{-}p$} (F);
 	\draw [tran,dashed,->,rounded corners] (B) -- +(0,.5) -| node[above,near start] {$1$}  (A);
 	\draw [tran,->,dashed,rounded corners] (F) -| node[below,near start] {$1$} (A);
 	\draw [tran,->,rounded corners] (F) -- +(.5,-.2) --  node[right] {$1$} +(.5,.2) -- (F);
 	\draw [tran,->,rounded corners] (C) -- +(.5,-.2) --  node[right] {$1$} +(.5,.2) -- (C);
 	\draw [tran,->,dashed,rounded corners] (A) -- +(-.5,-.2) --  node[left] {$1$} +(-.5,.2) -- (A);
 	\end{tikzpicture}
 	\caption{A simplified fdCTMC model of a communication protocol.}
 	\label{fig-exa-protocol}
 \end{figure}
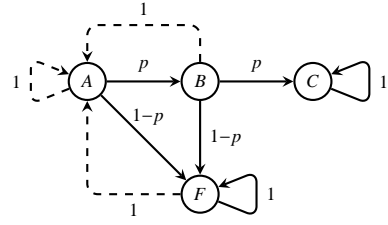

\begin{example}
\label{exa-protocol}
Consider a simple communication protocol where Alice tries to establish a connection with Bob via an unreliable communication channel. Alice starts by sending an \emph{Invite} message to Bob, and then she waits for Bob's \emph{Ack} message. Since each of these messages can be lost, Alice sets a timeout after which she restarts the protocol and sends another \emph{Invite} (the \emph{Ack} messages confirming a successful receipt of a ``previous'' \emph{Invite} are recognized and ignored). The protocol terminates when a connection is established, i.e., both messages are delivered successfully before the timeout. The behaviour of the unreliable channel is stochastic; a message is successfully delivered with a (known) probability~$p$, and the delivery time has a (known) distribution~$\mathit{Dtime}$. A simplified fdCTMC model of the protocol is given in Fig.~\ref{fig-exa-protocol}. The ``ordinary'' (i.e., exp-delay) and fixed-delay transitions are indicated by solid and dashed arrows, respectively, together with the associated probabilities. A faithful modeling of the $\mathit{Dtime}$ distribution using the phase-type approximation requires extra auxiliary states which are omitted\footnote{Hence, the simplified model corresponds to the situation when $\mathit{Dtime}$ is the exponential distribution with parameter~$\lambda$.} in Fig.~\ref{fig-exa-protocol} for the sake of simplicity (the main point is to illustrate the use of fixed-delay transitions). Note that the alarm clock is set in the initial state $A$, and it is switched off in the terminal state~$C$. If the alarm clock rings in any state except for $C$, the protocol is restarted and the alarm clock is reset. Now, the question is how to set the timeout so that the expected time needed to complete the protocol (i.e., to reach the state $C$ from the state $A$) is minimized. If the timeout is too large, a lot of time is wasted by waiting in the failure state~$F$. If it is too small, there is not enough time to complete the communication and the protocol is restarted many times before it succeeds. In this particular case, one may still  argue that an optimal timeout can be computed by hand and no synthesis algorithm is needed. Now consider a more complicated scenario where Alice tries to establish a simultaneous connection with $\text{Bob}_1,\ldots,\text{Bob}_n$ via different unreliable channels which are also unstable (i.e., an already established link with Bob$_i$ gets broken after a random time whose distribution is known). This scenario can still be modeled by a fdCTMC, and Alice needs to determine a suitable timeout which achieves the optimal expected time of completing the protocol. Since the properties of the individual channels can be different and the probability of breaking an already established connection increases as more and more Bobs get connected, the timeout chosen by Alice should actually depend on the subset of connections that remain to be established. The corresponding tuple of optimal timeouts is hard to compute manually. However, as we shall see in Section~\ref{sec:experiments}, a solution can be synthesized by our algorithm.  
\IEEEQEDopen
\end{example}
\smallskip

Now we explain the objective of fixed-delay synthesis. Intuitively, given a fdCTMC, the goal is to compute a delay function which minimizes the expected total cost incurred before reaching a given set of target states $G$ starting in a given initial state $\initstate$. For the fdCTMC of Example~\ref{exa-protocol}, the set of target states is $\{C\}$, the initial state is~$A$, and the costs correspond to the elapsed time. Our aim is to model general performance measures (not just the elapsed time), and therefore we use the standard cost structures that assign numerical costs to both states and transitions (see,~e.g.,~\cite{Puterman:book}). More precisely, we consider the following three cost functions: $\rateRew: \states \to \Rsetp$, which assigns a cost rate $\rateRew(s)$ to every state $s$ so that the cost $\rateRew(s)$ is paid for every time unit spent in the state~$s$, and functions $\impRewExp, \impRewFix: \states \times \states \to \Rsetpo$ that assign to each exp-delay and fixed-delay transition the associated execution cost. 

For every delay function $\timeouts$, let $\expred_{\timeouts}$ be the expected total cost incurred before reaching a target state of $G$ starting in the initial state $\initstate$ (note that when $\timeouts$ is fixed, the behaviour of the considered fdCTMC is fully probabilistic). For a given $\eps>0$, we say that a delay function $\timeouts$ is \emph{$\eps$-optimal} if 
\[
 \left\lvert \; \expred_{\timeouts} - \inf_{\timeouts'} \expred_{\timeouts'} \; \right\rvert
 \;\; < \;\; \eps.  
\]
Here, $\timeouts'$ ranges over all 
delay functions. The \emph{fixed-delay synthesis problem} for fdCTMC is to compute an $\eps$-optimal delay function (for a given $\eps>0$).

\subsection{Our algorithm for the fixed-delay synthesis problem}
\label{sec-our-alg}

For purposes of this subsection, we fix a fdCTMC $(S,\lambda,\prob,\allact,\trans)$, cost functions $\rateRew$, $\impRewExp$, $\impRewFix$, an initial state $\initstate$, and a set of target states~$G$. 

As we already mentioned, our fixed-delay synthesis algorithm for fdCTMC
is \emph{symbolic} in the sense that it avoids explicit constructions of
large action spaces and allows to safely disregard a large subsets of
actions that correspond to discretized timeout values. To explain what all
this means, we need to introduce some extra notions. Let $\Soff = S
\smallsetminus \allact$ be the set of all states where fixed-delay
transitions are disabled. Further, let $\Sset \subseteq \allact$ be the set
of all states where a timeout can be (re)set, i.e., $\Sset$ consists of all
$s \in \allact$ such that $s$ has an incoming exp-delay transition from a
state of $\Soff$, or an incoming fixed-delay transition (from any
state). For the fdCTMC of Fig.~\ref{fig-exa-protocol}, we have that
$\Soff = \{C\}$ and $\Sset = \{A\}$ (note the timeout is never set in the
states $B$ and $F$, although the ``alarm clock'' is turned on in these
states). Without restrictions, we assume that the initial state $\initstate$
and the each target state of $G$ belong to $\Soff \cup \Sset$ (otherwise, we
can trivially adjust the structure of our fdCTMC).

Now, let us fix a 
delay function $\timeouts$. If the execution of
our fdCTMC is initiated in a state $s \in \Soff \cup \Sset$ (for $s \in
\Sset$, the timeout is set to $\timeouts(s)$), then a state $s'$ such that either $s' \in \Soff$, or $s' \in \Sset$ and the timeout is (re)set in~$s'$, is visited with probability one. Note that the timeout is (re)set in $s' \in \Sset$ if the transition used to enter $s'$ is either fixed-delay (i.e., the alarm clock just rang and needs to be set again), or exp-delay and the previous state belongs to $\Soff$ (i.e., the alarm clock was off and needs to set on now). Hence, for every $s \in \Soff \cup \Sset$, we can define the probability distribution $\mtran(s,\timeouts)$ over $\Soff \cup \Sset$, where $\mtran(s,\timeouts)(s')$ is the probability that the first visited state satisfying the above condition is~$s'$.  At the moment, it is not yet clear how to compute/approximate the distribution $\mtran(s,\timeouts)$, but it is correctly defined. Further,
for every $s \in \Soff \cup \Sset$, let $\mcost(s,\timeouts)$ be the expected total cost incurred before reaching a state $s'$ satisfying the above condition (and starting in~$s$). Thus, we obtain a \emph{discrete-time} Markov chain $\mdp_{\timeouts}$ with the set of states $\Soff \cup \Sset$ where each
state $s$ is assigned the cost $\mcost(s,\timeouts)$. For the fdCTMC of Fig.~\ref{fig-exa-protocol}, the structure of $\mdp_{\timeouts}$ is shown
in Fig.~\ref{fig-exa-Md}. Here, $\kappa[\timeouts(A)]$ is the probability of
executing at least two exp-delay transitions in time $\timeouts(A)$. 
Note that $\mcost(C,\timeouts)$ is independent of~$\timeouts$.  

\begin{figure}[t]\centering
	\begin{tikzpicture}[x=1.5cm,y=1.5cm,font=\scriptsize]
	\node[label={$\mcost(A,\timeouts)$}] (A) at (0,0) [ran] {$A$};
	\node[label={$\mcost(C,\timeouts)$}] (C) at (1.5,0) [ran] {$C$};
	\draw (A) [tran,->] -- node[above] {$\kappa[\timeouts(A)] \cdot p^2$} (C); 
	\draw [tran,->,rounded corners] (C) -- +(.5,-.2) --  node[right] {$1$} +(.5,.2) -- (C);
	\draw [tran,->,rounded corners] (A) -- +(-.5,-.2) --  node[left] {$1-\kappa[\timeouts(A)] \cdot p^2$} +(-.5,.2) -- (A);
	\end{tikzpicture}
	\caption{The structure of $\mdp_{\timeouts}$ for the fdCTMC of Fig.~\ref{fig-exa-protocol}.}
	\label{fig-exa-Md}
\end{figure}

It is not hard to show that the Markov chain $\mdp_{\timeouts}$ faithfully mimics the behaviour of the considered fdCTMC for the delay function~$\timeouts$. More precisely, $\expred_{\timeouts}$ (i.e., the expected total cost incurred in our fdCTMC before reaching a target state of $G$ starting in $\initstate$) is equal to the expected total cost incurred in $\mdp_{\timeouts}$ before reaching a state of $G$ starting in $\initstate$. 
Since we do not aim at computing $\expred_{\timeouts}$ for a given $\timeouts$ but on synthesizing a suboptimal $\timeouts$, the Markov chain $\mdp_{\timeouts}$ does not appear very useful. However, $\mdp_{\timeouts}$ can be transformed into a \emph{discrete-time Markov decision process} $\mdp$ which serves this goal. Here we use the result of \cite{BKKNR:QEST2015} which, for a given $\eps > 0$ and every $s \in \allact$, allows construct a \emph{finite} set $\Tval(s)$ of discrete timeout values such that an $\eps$-optimal $\timeouts$ is guaranteed to exist even if $\timeouts(s)$ is restricted to $\Tval(s)$. For technical reasons, we also put $\Tval(s) = \{\infty\}$ for all $s \in \Soff$. 

Note that for every $s \in \Sset$, the distribution $\mtran(s,\timeouts)$ and the cost $\mcost(s,\timeouts)$ depends just of $\timeouts(s)$. To simplify our notation, we often write $\mtran(s,\tau)$ and $\mcost(s,\tau)$ to denote $\mtran(s,\timeouts)$ and $\mcost(s,\timeouts)$ where $\timeouts(s) = \tau$. For $s \in \Soff$, the distribution $\mtran(s,\timeouts)$ and the cost $\mcost(s,\timeouts)$ are independent of $\timeouts$. To unify our notation for all elements of $\Soff \cup \Sset$, we write $\mtran(s,\tau)$ and $\mcost(s,\tau)$ also when $s \in \Soff$, even if the $\tau$ is irrelevant. 

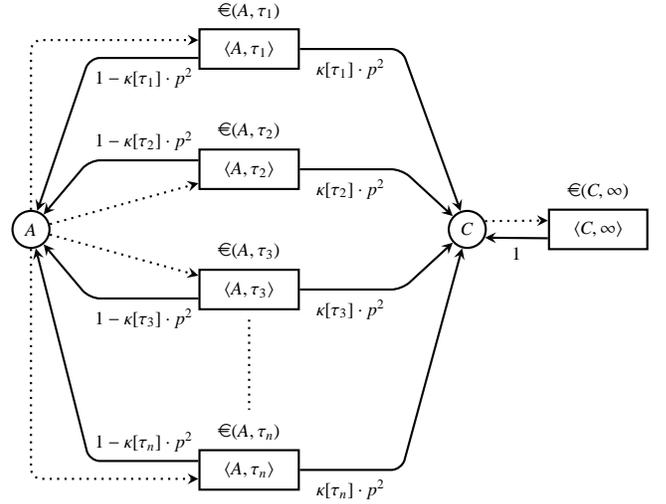
\begin{figure}[t]\centering
	\begin{tikzpicture}[x=2.9cm,y=.8cm,font=\scriptsize]
	\node (A) at (0,0) [ran] {$A$};
	\node (C) at (2,0) [ran] {$C$};
	\node[label={$\mcost(C,\infty)$}] (D) at (2.6,0) [act] {$~~\ac{C,\infty}~~$};
	\foreach \x/\y in {3/1,1/2,-1/3,-4/n}{
	     \node[label={$\mcost(A,\tau_{\y})$}] (A\y) at (1,\x) [act] {$~~\ac{A,\tau_{\y}}~~$};
	     \draw [tran, rounded corners, ->] (A\y) -- node[below] {$\kappa[\tau_{\y}] \cdot p^2$} +(.7,0) -- (C); 
    } 
    \draw [tran, rounded corners, ->] (A1.190) -- node[below] {$1-\kappa[\tau_{1}] \cdot p^2$} +(-.5,0) -- (A); 
    \draw [tran, rounded corners, ->] (A2.170) -- node[above] {$1-\kappa[\tau_{2}] \cdot p^2$} +(-.5,0) -- (A); 
    \draw [tran,dotted,->] (A) -- (A2); 
    \draw [tran,dotted,rounded corners,->] (A) |- (A1.170); 
    \draw [tran, rounded corners, ->] (A3.190) -- node[below] {$1-\kappa[\tau_{3}] \cdot p^2$} +(-.5,0) -- (A); 
    \draw [tran, rounded corners, ->] (An.170) -- node[above] {$1-\kappa[\tau_{n}] \cdot p^2$} +(-.5,0) -- (A); 
    \draw [tran,dotted,->] (A) -- (A3); 
    \draw [tran,dotted,rounded corners,->] (A) |- (An.190); 
    \draw [tran,dotted,-] (A3)+(0,-.5) -- +(0,-2); 
    \draw [tran,dotted,->] (C.25) -- (D.170); 
    \draw [tran,->] (D.190) --  node[below] {$1$} (C.335); 
	\end{tikzpicture}
	\caption{The structure of $\mdp$ for the fdCTMC of Fig.~\ref{fig-exa-protocol}.}
	\label{fig-exa-MDP}
\end{figure}

The MDP $\mdp$ is constructed as follows. For every state $s \in \Soff \cup \Sset$ and every $\tau \in \Tval(s)$, we add a special action $\ac{s,\tau}$ enabled in $s$. The outgoing transitions of $\ac{s,\tau}$ are now ``copied'' from $\mdp_{\timeouts}$, i.e., the probability of entering a state $s' \in \Soff \cup \Sset$ after selecting the action $\ac{s,\tau}$ is $\mtran(s,\tau)(s')$. Further, the action $\ac{s,\tau}$ is assigned the cost $\mcost(s,\tau)$. For the fdCTMC of Fig.~\ref{fig-exa-protocol}, the structure of $\mdp$ is shown in Fig.~\ref{fig-exa-MDP}.

An $\eps$-optimal delay function $\timeouts$ can now be obtained by computing an optimal stationary policy minimizing the expected total cost incurred in $\mdp$ before reaching a target state of $G$ starting in $\initstate$ (this can be achieved by a standard policy improvement algorithm; see, e.g., \cite{Puterman:book}). For every $s \in \Sset$, we put $\timeouts(s) = \tau_j$, where $\ac{s,\tau_j}$ is the action selected by the optimal stationary policy. For the remaining $s \in \allact \smallsetminus \Sset$, we set $\timeouts(s)$ arbitrarily.

The fixed-delay synthesis algorithm of \cite{BKKNR:QEST2015} constructs the MDP $\mdp$ explicitly, where all $\mtran(s,\tau)$ and all $\mcost(s,\tau)$ are approximated up to a sufficiently small error before computing an optimal policy. Note that for the fdCTMC of Fig.~\ref{fig-exa-protocol}, this essentially means to try out all possibilities in the discretized candidate set~$\Tval(A)$. Since the candidate sets $\Tval(s)$ are large, this approach cannot be applied to larger instances. 

The algorithm presented in this paper avoids the explicit construction of $\mdp$. The key idea is to express 
$\mtran(s,\tau)$ and $\mcost(s,\tau)$ \emph{analytically} as functions of~$\tau$. More precisely, for each $s \in \Soff \cup \Sset$, we consider the following two functions:
\begin{itemize}
	\item $\mtran_s : \Rsetpo \rightarrow \dist(\Soff {\cup} \Sset)$, where $\dist(\Soff {\cup} \Sset)$ is the set of all probability distributions over $\Soff \cup \Sset$. The function is defined by $\mtran_s(\tau) = \mtran(s,\tau)$.
	\item $\mcost_s : \Rsetpo \rightarrow \Rsetpo$ defined by $\mcost_s(\tau) = \mcost(s,\tau)$.
\end{itemize}
Further, for every $s,s' \in \Soff \cup \Sset$, let
\begin{itemize}
	\item $\mtran_{s,s'} : \Rsetpo \rightarrow \Rsetpo$
	be defined by $\mtran_{s,s'}(\tau) = \mtran(s,\tau)(s')$. 
\end{itemize}
The functions $\mtran_s$ and $\mcost_s$ can be expressed as certain infinite sums, but for every fixed error tolerance, this sum can be effectively truncated to finitely many summands. A precise definition is postponed to Section~\ref{sec-alg-full}. The functions $\mtran_s$ and $\mcost_s$ are then used in the symbolic policy improvement algorithm. We start with some (randomly chosen) eligible delay function such that $\timeouts(s) \in \Tval(s)$ for all $s \in \allact$. Then, we repeatedly improve $\timeouts$ until no progress is achieved. Each improvement round has two phases. First, we evaluate the \emph{current} $\timeouts$ in all states of $\Soff \cup \Sset$. That is, for each $s \in \Soff \cup \Sset$ we approximate the value $\expred_{\timeouts}^s$, which is equal to $\expred_{\timeouts}$ when the initial state is changed to $s$, up to a sufficient precision. Then, for each state $s \in \Sset$, we try to identify the action $\ac{s,\tau}$ such that the timeout $\tau$ \emph{minimizes} the function $K_s : \Rsetpo \rightarrow \Rsetpo$ defined by 
\[
   K_s(\tau) \quad = \quad \sum_{s' \in \Soff {\cup} \Sset} \mtran_{s,s'}(\tau) \cdot \expred_{\timeouts}^{s'} +  \mcost_s(\tau)
\] 
Instead of trying out all $\tau \in \Tval(s)$ one by one, we compute the \emph{symbolic derivative} of $K_s$, which is possible due to the analytical form of $\mtran_s$ and $\mcost_s$. Further, it turns out that the derivative takes zero value iff a certain effectively constructable \emph{univariate polynomial} takes zero value. Hence, we only need to deal with those $\tau \in \Tval(s)$ which are ``close'' to the roots of this polynomial, and we may safely ignore the others. Since the roots of univariate polynomials are easy to approximate using modern mathematical software (such as Maple), this approach is rather efficient and the set of relevant $\tau$'s obtained in this way is \emph{much} smaller than $\Tval(s)$. This is why our algorithm outperforms the one of \cite{BKKNR:QEST2015} so significantly.

\subsection{Related work}
\label{sec-related}

The relationship to the work of \cite{BKKNR:QEST2015} was already explained in Section~\ref{sec-our-alg}. In particular, we use the discretization constants developed in \cite{BKKNR:QEST2015} to define the sets $\Tval(s)$ (see Section~\ref{sec-our-alg}).  

Our fdCTMC formalism can be seen as a subclass of deterministic and
stochastic Petri nets~\cite{marsan1987petri}. The main restriction is that
in fdCTMC, at most one fixed-delay event can be enabled at a time (i.e., we
cannot have two different ``alarm clocks'' turned on simultaneously).
fdCTMC can also be seen as a special variant of Markov regenerative processes~\cite{ABD:MarkovRegenProcess-qest14}. Another related formalism are
delayed CTMC introduced in \cite{guet2012delayed}. Fixed-day events were used to model,
e.g., deterministic durations in train control systems \cite{Z:ECTS_synthesis}, time of server rejuvenation
\cite{german-book}, timeouts in power management systems \cite{QWP99}. Some of these models contain specific impulse or rate costs.

To the best of our knowledge, no generic framework for fixed-delay synthesis in stochastic continuous-time
systems has been developed so far. In previous works, some special cases were considered,
e.g., timeout synthesis in finite models~\cite{CRV:real-time-testing-synthesis,WDQ:real-time-test-execution}, history dependent timeouts~\cite{KNSS:PTA-scheduler-synthesis,JT:PTMDP-learning}, or timeout synthesis for a specific concrete model~\cite{XSCT:TRIVEDI_timeout_synthesis}. 

There is a number of papers on synthesizing other parameters of continuous-time
models, such as parametric timed automata~\cite{Alur-TA-params}, parametric
one-counter automata~\cite{hkow-concur09}, parametric Markov
models~\cite{DBLP:journals/sttt/HahnHZ11}, etc.  In the context of
continuous-time stochastic systems, the synthesis of appropriate rates in
CTMC was studied in \cite{Katoen-param-synt,CTMC-biochem,ceska}. In \cite{Katoen-param-synt}, a symbolic technique similar to ours is used to synthesize optimal rates in CTMC, but the results are not directly applicable in our setting due to the difference in objectives and possible cycles in the structure of fdCTMC. In \cite{TBR-in_CTMDP-Neuhausser,Prob_reach__param_DTMC,BKKKR:GSMP-games-TA,DBLP:journals/iandc/BrazdilFKKK13} 
the optimal controller synthesis for continuous-time (Semi)-Markov decision
processes is studied, which can be also seen as a synthesis problem for \emph{discrete} parameters in
continuous-time systems (contrary to our result, the schedulers are only
allowed to choose actions from a priori discrete and finite domains).

\section{The Algorithm}
\label{sec-algorithm}

In this section, we present our fixed-delay synthesis algorithm. In Section~\ref{sec-defs}, we give the technical details that were omitted in Sections~\ref{sec-fdCTMC} and~\ref{sec-our-alg}. Then, we continue with presenting the algorithm in Section~\ref{sec-alg-full}. We assume familiarity with basic notions of probability theory (such as probability space, random variable, expected value, Markov chain, Markov decision process) and with the notions introduced in  Sections~\ref{sec-fdCTMC} and~\ref{sec-our-alg}.

\subsection{Preliminaries}
\label{sec-defs}

We use $\Nset$, $\Nseto$, $\Rsetpo$, and $\Rsetp$ to denote the set of all positive integers, non-negative integers, non-negative real numbers, and positive real numbers, respectively. For a finite or countably infinite set $A$, we denote by $\dist(A)$ the set of all discrete probability distributions over $A$, i.e., functions $\mu: A \to \Rsetpo$ such that $\sum_{a\in A} \mu(a) = 1$. 

Recall (see Section~\ref{sec-fdCTMC}) that a fdCTMC is a tuple $\fdC = (S,\lambda,\prob,\allact,\trans)$, and that we use three cost functions \mbox{$\rateRew: \states \to \Rsetp$},  \mbox{$\impRewExp, \impRewFix: \states \times \states \to \Rsetpo$} to model performance measures. Also recall that $\Soff$ denotes the set $S \smallsetminus \allact$, and $\Sset$ denotes the set of all $s \in \allact$ such that $\prob(s',s) > 0$ for some $s' \in \Soff$, or $\trans(s',s) > 0$ for some $s' \in S$. 
A delay function is a function $\timeouts: \allact \to \Rsetp$. 
Further, we fix an initial state $\initstate \in \Soff \cup \Sset$ and a non-empty set of target states~$G \subseteq \Soff \cup \Sset$.

Now we formally define the semantics of fdCTMC. A \emph{configuration} is a pair $(\sta,\delays)$ where $\sta\in\states$ is the current state and $\delays \in \Rsetp \cup \{\nodel\}$ is the remaining time to perform a fixed-delay transition (i.e., the remaining time before the ``alarm clock'' rings). As we shall see, $d=\infty$ iff  $\sta\not\in \allact$. To simplify our notation, we extend each delay function $\timeouts$ also to the states of $\Soff$ by stipulating $\timeouts(\sta) = \infty$ for all $\sta \in \Soff$. A \emph{run} of $\fdC(\timeouts)$ starts in the configuration $(\sta_0,\delays_0)$ where $\sta_0 = \initstate$ and $\delays_0 = \timeouts(\initstate)$. If the current configuration of a run is $(\sta_i,\delays_i)$, then some random time $t_i$ is spent in $\sta_i$, and then a~next configuration $(s_{i+1},d_{i+1})$ is entered. Here, the time $t_i$ and the configuration $(s_{i+1},d_{i+1})$ are determined as follows: First, a random time $t_{exp}$ is chosen according to the exponential distribution with the rate $\lambda$. Then,
\begin{itemize}
	\item if $t_{exp}<d_i$, then an exp-delay transition is selected according to $\prob$, i.e., $t_i=t_{exp}$, $s_{i+1}$ is chosen randomly with probability $P(s_i,s_{i+1})$, and $\delays_{i+1}$ is determined as follows:
	\[
	  \delays_{i+1} = \begin{cases}
	                     \delays_i - t_{exp} & \text{if $\sta_{i+1} \in \allact$ and $\sta_i\in \allact$}\\
	                     \timeouts(\sta_{i+1}) & \text{if $\sta_{i+1} \not\in \allact$ or $\sta_i \not\in\allact$;} 
	  \end{cases}
	\]
	\item if $t_{exp} \geq d_i$, then a fixed-delay transition occurs, i.e., $t_i=d_i$, $s_{i+1}$ is chosen randomly with probability $\trans(s_i,s_{i+1})$, and $d_{i+1}=\timeouts(\sta_{i+1})$.
\end{itemize}
The corresponding probability space over all runs (i.e., infinite sequences of the form $(s_{0},d_{0}),t_0,(s_1,d_1),t_1,\dots$) is defined in the standard way (see, e.g., \cite{BKKNR:QEST2015}). We use $\probm_{\fdC(\timeouts)}$ to denote the associated probability measure. Further, we define a random variable $\costFdC$ assigning to each run $\ctrun = (\sta_0,\delays_0), t_0, (\sta_1,\delays_1), t_1,\ldots$ the \emph{total cost before reaching $\goalStates$} (in at least one transition), given by
\[
\costFdC(\ctrun) = \begin{cases}
                      \sum_{i=0}^{n-1} \left( t_i\cdot\rateRew(s_i) + \impRew_i(\ctrun) \right)
     & \text{for the least $n>0$}\\
     & \text{such that $s_n\in\goalStates$,} \\[1ex]
\infty & \text{if there is no such $n$,}
\end{cases}
\] 
where $\impRew_i(\ctrun)$ equals $\impRewExp(s_i,s_{i+1})$ for an exp-delay transition, i.e., when $t_i < \delays_i$, and equals $\impRewFix(s_i,s_{i+1})$ for a fixed-delay transition, i.e., when $t_i = \delays_i$. The expected value of $\costFdC$ (with respect to $\probm_{\fdC(\timeouts)}$) is denoted by $\expred_{\fdC(\timeouts)}$.

\subsection{A description of the algorithm}
\label{sec-alg-full}

For the rest of this section, we fix a fdCTMC $\fdC = (S,\lambda,\prob,\allact,\trans)$, cost functions $\rateRew$, $\impRewExp$, $\impRewFix$,
an initial state $\initstate \in \Soff \cup \Sset$, a non-empty set of target states~$G \subseteq \Soff \cup \Sset$, and $\eps > 0$. We assume that $\inf_{\timeouts'} \expred_{\fdC(\timeouts')} < \infty$, because the opposite case
can be easily detected (see~\cite{BKKNR:QEST2015}).

Due to \cite{BKKNR:QEST2015}, there effectively exist two positive rational numbers $\delta, \tau_{\max}$ such that if we put
\[
  \Tval(s) = \big\{k \cdot \delta \mid k\in\Nset 
  . \; k\cdot\delta \leq \tau_{\max} 
  \big\}
\]
for all $s \in \allact$, then there exists an $\eps$-optimal delay function $\timeouts$ which is \emph{$\Tval$-compatible}, i.e., $\timeouts(s) \in \Tval(s)$ for all $s \in \allact$. For technical reasons, we also stipulate $\Tval(s) = \{\infty\}$ for all $s \in \Soff$.

Now recall the MDP $\mdp$ introduced in Section~\ref{sec-our-alg}. We assume that all states of $\Soff \cup \Sset$ are reachable from $\initstate$, i.e., for every $s \in \Soff \cup \Sset$, there is a finite sequence $s_0,\ac{s_0,\tau_0},s_1,\ac{s_1,\tau_1},\ldots,s_n$ such that $s_0 = \initstate$, $s_n = s$, and 
$\mtran(s_i,\tau_i)(s_{i+1}) > 0$ for all $i < n$. Clearly, all states that are not reachable from $\initstate$ can be safely erased.

A \emph{policy} for $\mdp$ is a function $\sigma$ which to every $s \in \Soff \cup \Sset$ assigns an action $\ac{s,\tau}$ enabled in~$s$. Every policy $\sigma$ determines a unique probability space over all \emph{runs} in $\mdp$, i.e., infinite sequences of the form $\vtx_0, \ac{\vtx_0,\tau_0}, \vtx_1, \ac{\vtx_1,\tau_1},\ldots$ where $\vtx_0 = \initstate$. We use $\probm_{\mdp(\sigma)}$ to denote the associated probability measure. For each such run we also define the total cost incurred before reaching $\goalStates$ as $\sum_{i=0}^{n-1} \mcost(\vtx_{i},\ac{\vtx_i,\tau_i})$, where $n > 0$ is the least index such that $\vtx_n \in \goalStates$; if there is no such $n$, the total cost is set to $\infty$. The expected value of this cost with respect to  $\probm_{\mdp(\sigma)}$ is denoted by $\expred_{\mdp(\sigma)}$. We say that a policy $\sigma$ is \emph{optimal} if $\expred_{\mdp(\sigma)} =\min_{\sigma'} \expred_{\mdp(\sigma')}$, where $\sigma'$ ranges over all policies for~$\mdp$. We use $\expred_{\mdp[s](\sigma)}$ to denote $\expred_{\mdp(\sigma)}$ where the initial state is changed to~$s$.

Every policy $\sigma$ for $\mdp$ determines a $\Tval$-compatible delay function $\timeouts_\sigma$ given by $\timeouts_\sigma(s) = \tau$, where $\sigma(s) = \ac{s,\tau}$ and $s \in \Soff \cup \Sset$ (for $s \in \allact \smallsetminus \Sset$, the value of $\timeouts_\sigma(s)$ is irrelevant and it can be set to an arbitrary element of $\Tval(s)$). Conversely, every $\Tval$-compatible delay function $\timeouts$ determines a policy $\sigma_{\timeouts}$ in the natural way. Therefore, we do not formally distinguish between policies for $\mdp$ and $\Tval$-compatible delay functions (see, e.g., Algorithm~\ref{alg:pol-iter}).

Recall the distribution $\mtran(s,\tau)$ and the cost $\mcost(s,\tau)$ given in Section~\ref{sec-our-alg}. Now we give analytical definitions of the two crucial functions $\mtran_s : \Rsetpo \rightarrow \dist(\Soff {\cup} \Sset)$ and $\mcost_s : \Rsetpo \rightarrow \Rsetpo$ that have been introduced in Section~\ref{sec-our-alg}. For $s \in \Soff$, we simply put
\begin{align*}
   \mtran_s(\tau) & \quad = \quad  \prob(s, \cdot)\\
   \mcost_s(\tau) & \quad = \quad \frac{\rateRew(s)}{\lambda} + \sum_{s' \in S}\prob(s,s')\cdot\impRewExp(s,s')
\end{align*}
Observe that both functions are constant, and one can easily verify that both formulas agree with their definition, i.e., $\mtran_s(\tau) = \mtran(s,\tau)$ and 
$\mcost_s(\tau) = \mcost(s,\tau)$. Now let $s \in \Sset$. Then $\mtran_s(\tau)$ and $\mcost_s(\tau)$ need to 
``summarize'' the behaviour of our fdCTMC $\fdC$ starting in the configuration $(s,\tau)$ until a state $s'$ is reached such that either $s' \in \Soff$, or $s' \in \Sset$ and the timeout is reset in $s'$. Such a state can be reached after performing $i$ exp-delay transition, where $i$ ranges from zero to infinity. For each such $i$, we evaluate the probability of performing precisely $i$ exp-delay transitions before the timeout $\tau$ (using Poisson distribution), and then analyze the $i$-step behaviour in exp-delay transitions. To achieve that, we define the stochastic matrix $\overline{\prob} \in \Rsetpo^{S\times S}$ where $\overline{\prob}(s,\cdot) =  \prob(s,\cdot)$ for all $s \in \allact$, and  
$\overline{\prob}(s,s) = 1$ for all $s \in \Soff$. In other words, $\overline{\prob}$ is the same as $\prob$ but all states of $\Soff$ are now absorbing. Further, we use $\vec{1}_s$ to denote a row vector such that $\vec{1}_s(s) = 1$ and $\vec{1}_s(s') = 0$ for all $s'\neq s$. Thus, we obtain
\begin{align*}
\mtran_s(\tau) & = \sum_{i=0}^{\infty} e^{-\lambda\tau}\frac{(\lambda\tau)^i}{i!} \cdot \left( \vec{1}_s \cdot \overline{\prob}^i \right) \cdot \trans 
\end{align*}
The function $\mcost_s(\tau)$ is slightly more complicated, because we also need to evaluate the total costs incurred before reaching a state $s'$ satisfying the condition stated above. Here we also employ a function 
$\overline{\rateRew}$ which is the same as $\rateRew$ but returns~$0$ for all states of $\Soff$, and functions $\overline{\expectedRewExp},\overline{\expectedRewFix}: \states \to \Rsetpo$ that assign to each state the expected impulse cost of the next exp-delay and the next fixed-delay transition, respectively.

\begin{align*}
\mcost_s(\tau)  =  \sum_{i=0}^{\infty}
e^{-\lambda\tau}\frac{(\lambda\tau)^i}{i!} 
& \left( ~
\sum_{j=0}^{i-1} \left(\vec{1}_s \cdot \overline{\prob}^j\right) \cdot
\left(\frac{\tau\cdot\overline{\rateRew}}{i+1} + \overline{\expectedRewExp}
\right) \right. 
\\ 
& \left. ~~  \;+\;
\left(\vec{1}_s \cdot \overline{\prob}^i\right) \cdot
\left(\frac{\tau\cdot\overline{\rateRew}}{i+1} +
  \overline{\expectedRewFix}\right) ~
\right) 
\end{align*}

\noindent
Again, one can verify that $\mtran_s(\tau) = \mtran(s,\tau)$ and  $\mcost_s(\tau) = \mcost(s,\tau)$. For more detailed explanation and proof please refer to \cite{BKKNR:new-arxiv}.

Since the series $\mtran_s(\tau)$ and $\mcost_s(\tau)$ are defined by
infinite sums, the next step is to compute a large enough $I \in \Nset$ such
that the first $I$ summands of $\mtran_s(\tau)$ and $\mcost_s(\tau)$
approximate $\mtran(s,\tau)$ and $\mcost(s,\tau)$ with a sufficient
accuracy. Here we borrow another result of \cite{BKKNR:QEST2015}, where a
sufficiently small approximation error $\kappa$ for evaluating
$\mtran(s,\tau)$ and $\mcost(s,\tau)$ when constructing the MDP $\mdp$ was
given. Hence, it suffices to find a sufficiently large $I \in \Nset$ such
that the reminder of the constructed series is bounded by $\kappa$ (for all
$s \in \Soff \cup \Sset$ and $\tau \in \Tval(s)$). Since we have an upper
bound $\tau_{\max}$ on the size of $\tau$, an appropriate $I$ can be
computed easily. From now on, we use $\mtran_s^I(\tau)$ and
$\mcost_s^I(\tau)$ as  $\mtran_s(\tau)$ and $\mcost_s(\tau)$, respectively,
where the infinite sums are truncated to the first $I$
summands only.

\begin{algorithm}[t] 
\SetAlgoLined
\DontPrintSemicolon
\SetKwInOut{Input}{input}\SetKwInOut{Output}{output}
\SetKwData{n}{n}\SetKwData{f}{f}\SetKwData{g}{g}
\SetKwData{Low}{l}\SetKwData{x}{x}

\Input{$\mdp$ and a $\Tval$-consistent delay function $ \timeouts' $}
\Output{a $\Tval$-consistent delay function $\timeouts$ optimal for $\mdp$}
\BlankLine
\Repeat{$\timeouts$ = $\timeouts'$} {
  $ \timeouts := \timeouts' $ \; \tcp{\strategyEvaluation} 
  Compute a vector $ \valueVector $ such that $\valueVector(s) := \expred_{\mdp[s](\timeouts)}
  $ \label{alg:strategy-eval} \;
\ForEach{ $ \sta \in \Sset $} {    
\tcp{\strategyImprovement} 
$ \candidates := \argmin_{\tau
      \in \Tval(s)}{ \mtran^I_\sta(\tau) \cdot
      \valueVector  + \mcost^I_\sta(\tau)
    }$ \label{alg:strategy-improvement} \; 
    \eIf {$\timeouts(\sta) \in \candidates $} {
      $ \timeouts'(\sta) := \timeouts(\sta) $ \;
    } {
      $ \timeouts'(\sta) := \min \candidates $
    } 
  }
}
\caption{Policy Iteration for $\mdp$~\cite{Puterman:book}}
\label{alg:pol-iter}
\end{algorithm}

As we already mentioned, our fixed-delay synthesis algorithm is essentially a ``symbolic'' variant of the standard policy iteration algorithm \cite{Puterman:book} applied to the MDP $\mdp$ where the actions of $\mdp$ are not constructed explicitly but generated ``on demand''. We start by recalling the standard policy iteration which assumes that $\mdp$ is given explicitly (see Algorithm~\ref{alg:pol-iter}). This algorithm starts with some (arbitrary) $\Tval$-consistent delay function $\timeouts'$ (recall that we do not distinguish between $\Tval$-consistent delay functions and policies) and gradually improves this function until reaching a fixed point. Each iteration consists of two phases: \emph{\strategyEvaluation{}} and \emph{\strategyImprovement}. In the \strategyEvaluation{} phase, the vector $\valueVector$ is computed, such that  $\valueVector(s)$
is the expected total cost until reaching a target state when starting from
$\sta$ and using the policy $\timeouts$. This can be done in polynomial time by
solving a set of linear equations. In the
\strategyImprovement{} phase, 
a new delay function is obtained by choosing a new action separately for each
state of $\Sset$\footnote{For the remaining states we have only one action, so there is nothing to improve.}. First, the set of actions $\argmin_{\tau\in\Tval(s)}{ \mtran^I_\sta(\tau)
  \cdot \valueVector + \mcost^I_\sta(\tau) } $ is computed and then some of them is picked, but the old action $
\timeouts(\sta) $ must be chosen whenever possible. Policy iteration
terminates 
in a finite number of steps and returns an optimal
\strategy~\cite{F:exp-pol-iter}. 

Our symbolic algorithm is obtained by modifying Algorithm~\ref{alg:pol-iter}. The policy evaluation step is efficient and here we do not need to implement any changes. In the policy improvement, we proceed differently. Due to our analytical representation of $\mtran_s^I(\tau)$ and $\mcost_s^I(\tau)$, we can now interpret $\mtran^I_\sta(\tau) \cdot \valueVector  + \mcost^I_\sta(\tau)$ as a function $\func$ of the variable~$\tau$. Now we show that $\func$ has a nice property---it is a continuous expolynomial function.

\begin{proposition} \label{prop:polynom}
For all $\sta \in \Sset$ and  $\valueVector \in \Rsetpo^{|\Soff \cup \Sset|} $, 
we have that 
$$\func(\tau) = e^{-\lambda\tau} \cdot \pol(\tau),$$
where $\pol(\tau)$ is a univariate polynomial whose degree is bounded by~$I$.
\end{proposition}
Proposition~\ref{prop:polynom} follows directly from the definition of $\mtran_s^I(\tau)$ and 
$\mcost_s^I(\tau)$.
Note that $ \func $  is continuous and easily differentiable.
Hence, we can identify the (global) minima of $\func$ in the interval $[\alpha_s,\beta_s]$, where $\alpha_s = \min \Tval(s)$ and $\beta_s = \max \Tval(s)$, which are the points where the first derivative of $\func$ is zero, or the bounds of the interval.
Let $\func'$ be the first derivative of $\func$.  
Then 
\begin{align*}
 \func'(\tau) & = e^{-\lambda \tau} \cdot \big( \pol(\tau) \big)' + \big(e^{-\lambda \tau}\big)' \cdot \pol(\tau)\\
 & = e^{-\lambda \tau} \cdot \Big ( \big (\pol(\tau) \big )' - \lambda
\cdot \pol(\tau) \Big)
\end{align*}
where $e^{-\lambda \tau} >0$ 
for all $\tau \in \Rsetpo$.  Thus, we can restrict ourselves to root
isolation of a univariate polynomial 
\[ 
 \poll(\tau) = (\pol(\tau))' - \lambda \cdot \pol(\tau)
\]
with a finite degree bounded by $I$.  Note that if the polynomial has no real roots, the minimum of $\func$ on $[\alpha_s,\beta_s]$ is in the bounds of the interval. If there are infinitely many
roots, then $\func$ is a constant function and its minimum is realized by any~$\tau \in [\alpha_s,\beta_s]$.
Otherwise, there are at most $I$ real roots. 
Hence, it suffices to evaluate $\mtran_s^I$ and $\mcost_s^I$ in the bounds of the interval, and in all values of $\Tval(s)$ whose distance from the roots of $\poll$ is at most $\delta$, where $\delta$ is the discretization constant used to define $\Tval(s)$. This reasonably bounds
the number of $\mtran_s^I$ and $\mcost_s^I$ evaluations in each
\strategyImprovement{} step.

There are many tools and libraries that can efficiently isolate real roots
for large degree univariate polynomials to high precision. In our experiments, we used
Maple~\cite{maple}. The real roots for the largest generated polynomial (of degree $226$) were isolated with
precision $20$ decimal digits in $0.14$ seconds; see Section~\ref{sec:experiments} for more details.

\begin{algorithm}[t] 
\SetAlgoLined
\DontPrintSemicolon
\SetKwInOut{Input}{input}\SetKwInOut{Output}{output}
\SetKwData{n}{n}\SetKwData{f}{f}\SetKwData{g}{g}
\SetKwData{Low}{l}\SetKwData{x}{x}

\Input{A fdCTMC $\fdC$ and approximation error $\eps >0$ }
\Output{delay function $\timeouts$ that is $\eps$-optimal in $\expred_{\fdC(\timeouts)}$}
\BlankLine
Compute $\delta$, $\tau_{\max}$, $I$, and the states of $\mdp$
\; 
$\timeouts'(\sta) := \alpha_s $ for all $\sta \in \Sset$, and $\timeouts'(\sta) := \infty$ for all $\sta \in \Soff$ \; \label{alg:initd} 
\Repeat{$\timeouts$ = $\timeouts'$} {
  $ \timeouts := \timeouts' $ \;
  Compute a vector $ \valueVector $ such that $\valueVector(s) := \expred_{\mdp[s](\timeouts)}
   $ \label{alg:policy-eval-new} \; 
\ForEach{ $ \sta \in \Sset$ } {
    Compute the polynomial $\poll$\; 
    \eIf {$\poll=0$} {
      $\candidates := \Tval(s) $\; \label{alg:zero-degree} 
    } {
      Isolate all real roots $\roots$ of $\poll$ for accuracy $\delta/2$ \;
      
      $\paramvalspace'(\sta) :=  \{\alpha_s, \beta_s\} \cup 
          \bigcup_{r \in \roots} (\Tval(s) \cap [r- 3\cdot\delta/2 , r+ 3\cdot\delta/2 ])$ \; 
      Compute the set $ \candidates := \argmin_{\tau \in \paramvalspace'(\sta)}{ \mtran^I_\sta(\tau) \cdot \valueVector  + \mcost^I_\sta(\tau) }$ \label{alg:candidates}\;
    }
    \eIf {$\timeouts(\sta) \in \candidates $} {
      $ \timeouts'(\sta) := \timeouts(\sta) $ \;
    } {
      $ \timeouts'(\sta) := \min \candidates $ \label{alg:unambiguity}
    } \label{alg:strategy-improvement-new}
  }
}
\caption{Symbolic Policy Iteration for fdCTMC}
\label{alg:dsc-pol-iter-new}
\end{algorithm}

The pseudo-code of the resulting algorithm is given as Algorithm~\ref{alg:dsc-pol-iter-new}. First, we compute the constants $\delta$, $\tau_{\max}$, $I$, and the states of $\mdp$. 
Then we apply the policy iteration algorithm initiated to a delay vector of minimal delays $\alpha_s$.
We use the observations above to reduce the number of evaluations of $\mtran_s^I$ and $\mcost_s^I$ in the \strategyImprovement{} step. At line~\ref{alg:zero-degree} we know that $\func$ is constant, thus we assign the whole $\Tval(s)$ to the set $\candidates$ of all minimizing arguments. Otherwise we isolate the roots of $\poll$ and generate a set $\paramvalspace'(\sta)$ of all candidates for evaluation. 
Observe that if we would isolate the roots of $\poll$ exactly, we need to
evaluate both closest points from $\Tval(s)$, i.e., add
into $\paramvalspace'(\sta)$ all points of
$\Tval(s)$ in distance at most $ \delta $ from each of the
roots. Since we isolate the roots of $ \poll $ with accuracy $ \delta/2 $,
we need to add all numbers from $ \Tval(s)$ in distance
$ 3 \cdot \delta/2 $. Thus in each \strategyImprovement{} step we evaluate $\mtran_s^I$ and $\mcost_s^I$ for at
most $ I \cdot 4 + 2 $ numbers from $ \Tval(s)$ instead
of the whole $ \Tval(s)$. Now we can state and prove the correctness of our algorithm.

\begin{theorem} \label{thm:alg-correctness}
  Algorithm~\ref{alg:dsc-pol-iter-new} returns an $\eps$-optimal delay function.
\end{theorem}
\begin{IEEEproof} 
	Algorithm~\ref{alg:pol-iter} terminates and returns an optimal policy $\timeouts$ for an arbitrary initial policy \cite{F:exp-pol-iter}, and $\timeouts$ is an \mbox{$\eps$-optimal} delay function for $\fdC$ \cite{BKKNR:QEST2015}. Assume that Algorithm~\ref{alg:pol-iter} is given the initial  \strategy{} $\timeouts'$ of Algorithm~\ref{alg:dsc-pol-iter-new}, where  $\timeouts'(\sta) := \alpha_s $ for all $\sta \in \Sset$, and $\timeouts'(\sta) := \infty$ for all $\sta \in \Soff$. We show that then 
    Algorithm~\ref{alg:dsc-pol-iter-new} closely mimics Algorithm~\ref{alg:pol-iter}, i.e., after each \strategyImprovement{} step, both algorithms compute the same \delayVector{}s. Thus, we obtain that Algorithm~\ref{alg:dsc-pol-iter-new} returns an $\eps$-optimal \delayVector{} for $\fdC$. 
	
	For the sake of contradiction, assume that there is some $i \in \Nset$ such that the delay functions $\timeouts'$ produced by Algorithms~\ref{alg:pol-iter} and~\ref{alg:dsc-pol-iter-new} after $i$ iterations of the \strategy{} improvement loop are different, and let us further suppose that $i$ is the least index with this property. Hence, both algorithms start the $i$-th iteration of the strategy improvement loop with the same value stored in~$\timeouts'$.
	
	Observe that both algorithms work with the same action spaces $\Tval(s)$, and also the \strategyEvaluation{} steps are the same. Hence, both algorithms produce the same vector $\valueVector$.  The only difference is in the \strategyImprovement{} step, where Algorithm~\ref{alg:dsc-pol-iter-new} and Algorithm~\ref{alg:pol-iter} may choose different \strategies. 
	Let $\sta \in \Soff \cup \Sset$ be the first state where the $\timeouts'$ produced by the two algorithms differ, and let $\tau_{\ref{alg:dsc-pol-iter-new}}$ and $\tau_{\ref{alg:pol-iter}}$ be the $\timeouts'(\sta)$ of Algorithm~\ref{alg:dsc-pol-iter-new} and Algorithm~\ref{alg:pol-iter}, respectively. 
	Then there are four possibilities. 
	\begin{enumerate}
		\item $\func(\tau_{\ref{alg:dsc-pol-iter-new}}) < \func(\tau_{\ref{alg:pol-iter}})$, \label{enum:case1}
		\item $\func(\tau_{\ref{alg:dsc-pol-iter-new}}) = \func(\tau_{\ref{alg:pol-iter}})$ and $\tau_{\ref{alg:dsc-pol-iter-new}} < \tau_{\ref{alg:pol-iter}} $,  \label{enum:case2}
		\item $\func(\tau_{\ref{alg:dsc-pol-iter-new}}) > \func(\tau_{\ref{alg:pol-iter}})$,  \label{enum:case3}
		\item $\func(\tau_{\ref{alg:dsc-pol-iter-new}}) = \func(\tau_{\ref{alg:pol-iter}})$ and $\tau_{\ref{alg:dsc-pol-iter-new}} > \tau_{\ref{alg:pol-iter}} $. \label{enum:case4}
	\end{enumerate}
	Case~\ref{enum:case1} contradicts the minimality of $\func(\tau_{\ref{alg:pol-iter}})$ according to Algorithm~\ref{alg:pol-iter}. 
	Case~\ref{enum:case2} contradicts the minimality of $\tau_{\ref{alg:pol-iter}}$ according to Algorithm~\ref{alg:pol-iter} because 
	$\func(\tau_{\ref{alg:dsc-pol-iter-new}}) = \func(\tau_{\ref{alg:pol-iter}})$ and $\func(\tau_{\ref{alg:pol-iter}})$ is the minimum of $\{\func(\tau) \mid \tau \in \Tval(s) \}$ according to Algorithm~\ref{alg:pol-iter}.
	
	Now assume that Case~\ref{enum:case3} or Case~\ref{enum:case4} holds. We take the minimal $\tau_{\ref{alg:pol-iter}}$ of $\func$ according to Algorithm~\ref{alg:pol-iter} and show that Algorithm~\ref{alg:dsc-pol-iter-new} could not choose $\tau_{\ref{alg:dsc-pol-iter-new}}$. If $\tau_{\ref{alg:pol-iter}} = \alpha_s$ or $\tau_{\ref{alg:pol-iter}}= \beta_s$, then this claim is trivial.
	Otherwise, we find the closest local minimum of $\func$ that is $\leq \tau_{\ref{alg:pol-iter}}$ or $\geq \tau_{\ref{alg:pol-iter}}$ and denote it by $b_{\ref{alg:pol-iter}}$ and $a_{\ref{alg:pol-iter}}$, respectively. From the continuity of $\func$ we have that either
	\begin{itemize}
		\item    $\func(b_{\ref{alg:pol-iter}}) \leq \func(\tau_{\ref{alg:pol-iter}})$ and $\forall b \in [b_{\ref{alg:pol-iter}}, \tau_{\ref{alg:pol-iter}}]. \func(b_{\ref{alg:pol-iter}}) \leq \func(b) \leq \func(\tau_{\ref{alg:pol-iter}})$, or
		\item  $\func(a_{\ref{alg:pol-iter}}) \leq \func(\tau_{\ref{alg:pol-iter}})$ and $\forall a \in [\tau_{\ref{alg:pol-iter}}, a_{\ref{alg:pol-iter}}]. \func(a_{\ref{alg:pol-iter}}) \leq \func(a) \leq \func(\tau_{\ref{alg:pol-iter}})$,
	\end{itemize}
	i.e., there is a local minimum $ x \in \{b_{\ref{alg:pol-iter}}, a_{\ref{alg:pol-iter}} \}$ in $\func$ such that all values of $\func$ between $x$ and $\tau_{\ref{alg:pol-iter}}$ are smaller or equal to $\func(\tau_{\ref{alg:pol-iter}})$.
	
	We derive the contradiction just for the first case because the second one is 
	symmetric.
	The set $\Tval(s) \cap [b_{\ref{alg:pol-iter}},\tau_{\ref{alg:pol-iter}}]$ is non-empty since it contains at least $\tau_{\ref{alg:pol-iter}}$. Then the minimal number $m$ from set $\Tval(s) \cap [b_{\ref{alg:pol-iter}},\tau_{\ref{alg:pol-iter}}]$
	according to $\func$
	was clearly in $\paramvalspace'(\sta)$ since we made at least $\delta/2$ error when finding the root and we considered all numbers from $\Tval(S)$ within $1.5 \cdot \delta$ distance. 
	If Case~\ref{enum:case3} holds, then clearly $\func(m) \leq \func(\tau_{\ref{alg:pol-iter}}) < \func(\tau_{\ref{alg:dsc-pol-iter-new}})$ which contradicts the minimality of $\func(\tau_{\ref{alg:dsc-pol-iter-new}})$ according to Algorithm~\ref{alg:dsc-pol-iter-new} because $\func(m)$ would be chosen as the minimum.
	
	If Case~\ref{enum:case4} holds, then either $\func(m) < \func(\tau_{\ref{alg:pol-iter}}) = \func(\tau_{\ref{alg:dsc-pol-iter-new}})$ which contradicts the minimality of $\func(\tau_{\ref{alg:dsc-pol-iter-new}})$ according to Algorithm~\ref{alg:dsc-pol-iter-new}, or $\func(m) = \func(\tau_{\ref{alg:pol-iter}}) = \func(\tau_{\ref{alg:dsc-pol-iter-new}})$. In the latter case, either the polynomial has zero degree and Algorithm~\ref{alg:dsc-pol-iter-new} selects $\alpha_s$ (which is a contradiction since $\alpha_s \leq \tau_{\ref{alg:pol-iter}}$), or $m=\tau_{\ref{alg:pol-iter}}$ and this  contradicts the minimality of $\tau_{\ref{alg:dsc-pol-iter-new}}$ according to Algorithm~\ref{alg:dsc-pol-iter-new}.
\end{IEEEproof}

\section{Experimental Evaluation}\label{sec:experiments}

\begin{table*}[t]
\begin{center}
\begin{tabular}{| c | r | r | r | r | r | r |}
\hline
Num. of Bobs &  \multicolumn{1}{c |}{T1} & \multicolumn{1}{c |}{T2} & \multicolumn{1}{c |}{T3} & \multicolumn{1}{c |}{T4} & \multicolumn{1}{c |}{T5} & \multicolumn{1}{c |}{T6} \\
\hline 
1 & 3.779370 & & & & & \\
2 & 3.737017 & 3.868655 & & & & \\
3 & 3.661561 & 3.784139 & 3.946357 & & & \\
4 & 3.577685	& 3.684519 & 3.826398 & 4.014022 & & \\
5 & 3.498647 & 3.587113 & 3.705449 & 3.864535 & 4.073141 & \\
6 & 3.430744 & 3.501000 & 3.596000 & 3.724862 & 3.899238 & 4.125076 \\

\hline
\end{tabular}

\end{center}
\caption{The synthesized timeouts for Model I.}
\label{tab:sync2}
\end{table*}

In this section we present the results achieved by our ``symbolic'' Algorithm~\ref{alg:dsc-pol-iter-new}, and compare its efficiency against the ``explicit'' algorithm of \cite{BKKNR:QEST2015} and its outcomes that have been reported in  \cite{KRF:iFM2016}.

We start with some notes on implementation, and then compare the two algorithms on selected models.

\paragraph{The ``explicit'' algorithm of \cite{BKKNR:QEST2015}}
 
The implementation details of the algorithm are explained in \cite{KRF:iFM2016}. It is an extension of PRISM model checker \cite{KNP:prismCAV11} employing the explicit computation engine.  First, a finite discretized MDP is built using the optimizations reported in \cite{KRF:iFM2016}, and then this MDP is solved by the standard algorithms of PRISM. Currently there are three solution methods available for computing an optimal \strategy{} for total reachability cost in a finite MDP: policy iteration, value iteration, and Gauss-Seidl value iteration. The policy iteration has been identified as the fastest one.

\paragraph{The ``symbolic'' Algorithm~\ref{alg:dsc-pol-iter-new}}
We have a prototype implementation of Algorithm~\ref{alg:dsc-pol-iter-new} that is also implemented as an extension of PRISM and uses the ``symbolic'' policy iteration method. We tested several libraries and tools for isolating real roots of polynomials (Apache Commons, Matlab, Maple, and Sage). The best performance was achieved by Maple \cite{maple}, and we decided to use this software in our proof-of-concept implementation. 
Currently, we call Maple directly from Java, providing the polynomial and the required precision for the roots. We measure the CPU time for all Maple calls and add it to the final result. 

All the computations were run on platform HP DL980 G7 with 8 64-bit processors
Intel Xeon X7560 2.26GHz (together 64 cores) and 448 GiB DDR3 RAM.
The time and space was measured by the Linux command \code{time}. The $\text{N/A}$
result stands for out of memory exception.

\subsection{Model I., Communication protocol}

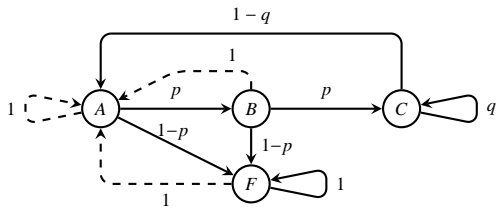
\begin{figure}[t]\centering
	\begin{tikzpicture}[x=2cm,y=1.cm,font=\scriptsize]
	\node (A) at (0,0)   [ran] {$A$};
	\node (B) at (1,0)   [ran] {$B$};
	\node (C) at (2,0)   [ran] {$C$};
	\node (F) at (1,-1)  [ran] {$F$};
	\draw [tran,->] (A) -- node[above] {$p$} (B);
	\draw [tran,->] (A) -- node[right, near start] {$1{-}p$} (F);
	\draw [tran,->] (B) -- node[above] {$p$} (C);
	\draw [tran,->] (B) -- node[right] {$1{-}p$} (F);
	\draw [tran,dashed,->,rounded corners] (B) -- +(0,.5) -- node[above,near start] {$1$}  +(-.5,.5) --  (A);
	\draw [tran,->,dashed,rounded corners] (F) -| node[below,near start] {$1$} (A);
	\draw [tran,->,rounded corners] (F) -- +(.5,-.2) --  node[right] {$1$} +(.5,.2) -- (F);
	\draw [tran,->,rounded corners] (C) -- +(.5,-.2) --  node[right] {$q$} +(.5,.2) -- (C);
	\draw [tran,->,rounded corners] (C) -- +(0,1) -|  node[above, near start] {$1-q$}   (A);
	\draw [tran,->,dashed,rounded corners] (A) -- +(-.5,-.2) --  node[left] {$1$} +(-.5,.2) -- (A);
	\end{tikzpicture}
	\caption{A fdCTMC model of the communication with Bob$_i$.}
	\label{fig-exa-protocol2}
\end{figure}

We start with the model discussed in Example~\ref{exa-protocol} where Alice is communicating with $\text{Bob}_1,\ldots,\text{Bob}_n$. The communication with $\text{Bob}_i$ is modeled as the fdCTMC of Fig.~\ref{fig-exa-protocol2}. So, the only difference from the fdCTMC of Fig.~\ref{fig-exa-protocol} is that now we also model the possibility of ``breaking'' an already established connection. We set $p = q = 0.9$, the rate costs are equal to $1$, all fixed-delay transition incur the impulse cost~$1$, and the exp-delay transitions incur zero cost. 

The whole protocol is modeled as a fdCTMC obtained by constructing the ``parallel composition'' of $n$ identical copies of the fdCTMC of Fig.~\ref{fig-exa-protocol2} (i.e., we assume that all Bobs use the same type of communication channel). The current state of this parallel composition is given by the $n$-tuple of current states of all components. In particular, the initial state is $(A,\ldots,A)$, and the only target state is $(C,\ldots,C)$. Obviously, the number of states grows exponentially with~$n$. 

Table~\ref{tab:sync} shows the outcomes achieved by the ``explicit'' and the ``symbolic'' algorithm. The first column gives the number of Bobs involved in the protocol, the second column is the error $\eps$, the third columns specifies the total number of states of the resulting fdCTMC model, the fourth and the fifth column specify the maximal number of roots and the maximal degree of the constructed polynomials in the ``symbolic'' algorithm,  and the last two columns give the time needed to compute the results. Note that the ``explicit'' algorithm cannot analyze a protocol with more that three Bobs, and tends to be significantly worse especially for smaller $\eps$.

\begin{table}
\begin{center}
\begin{tabular}{| c | c | r | r | r | r | r |}
\hline
\multicolumn{1}{|c|}{Num.} & \multirow{2}{*}{$\eps$}  & \multicolumn{1}{c|}{Num.} & \multicolumn{1}{c |}{Num.} & \multicolumn{1}{c|}{Max pol.} & \multicolumn{2}{c|}{CPU time [s]} \\
\cline{6-7}
\multicolumn{1}{|c|}{of Bobs} & & \multicolumn{1}{c|}{states} & \multicolumn{1}{c|}{roots} & \multicolumn{1}{c|}{degree} & \multicolumn{1}{c|}{symbolic} & \multicolumn{1}{c|}{explicit}  \\
\hline
1 & $10^{-2}$ & 4 & 8 & 55 & 2.91 & 4.4 \\
1 & $10^{-3}$ & 4 & 8 & 60 & 2.94 & 11.84 \\
1 & $10^{-4}$ & 4 & 8 & 64 & 2.96 & 75.18 \\
1 & $10^{-5}$ & 4 & 10 & 69 & 3.01 & 3429.88 \\
2 & $10^{-2}$ & 32 & 16 & 122 & 3.65 & 33.00 \\
2 & $10^{-3}$ & 32 & 20 & 129 & 4.93 & 1265.45 \\
2 & $10^{-4}$ & 32 & 20 & 135 & 4.91 & N/A \\ 
3 & $10^{-2}$ & 192 & 30 & 202 & 6.02 & 1765.71 \\ 
3 & $10^{-3}$ & 192 & 31 & 210 & 7.16 & N/A \\ 
3 & $10^{-4}$ & 192 & 32 & 220 & 7.47 & N/A \\
4 & $10^{-2}$ & 1024 & 40 & 280 & 10.71 & N/A \\
4 & $10^{-3}$ & 1024 & 40 & 290 & 11.41 & N/A \\
5 & $10^{-2}$ & 5120 & 55 & 360 & 26.36 & N/A \\
6 & $10^{-2}$ & 24576 & 65 & 449 & 221.76 & N/A \\

\hline
\end{tabular}
                 
\end{center}
\caption{Performance characteristics for Model~I.}
\label{tab:sync}
\end{table}

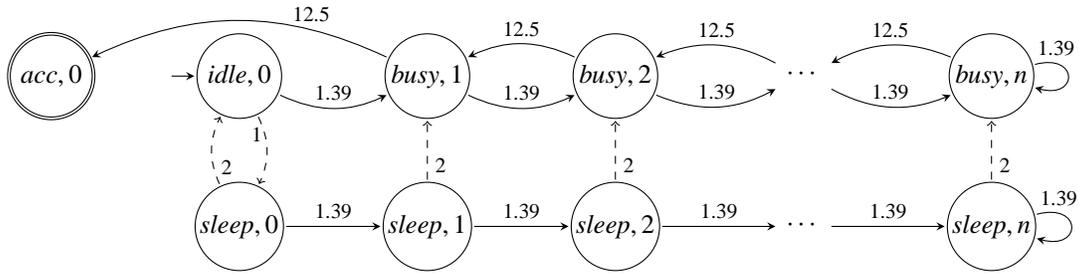
\begin{figure*}
\newcommand{\bend}{-25}
  \begin{center}
 \begin{tikzpicture}[outer sep=0.1em, xscale=1, yscale=1]
      \tikzstyle{fixed}=[dashed,->]; 
     \tikzstyle{fixed label}=[font=\small];
      \tikzstyle{exp}=[->,rounded corners,,>=stealth]; \tikzstyle{exp rate}=[font=\small];
      \tikzstyle{loc}=[draw,circle, minimum size=3.2em,inner sep=0.1em];
      \tikzstyle{accepting}+=[outer sep=0.1em]; \tikzstyle{loc
        cost}=[draw,rectangle,inner sep=0.07em,above=6, minimum
      width=0.8em,minimum height=0.8em,fill=white,font=\footnotesize];
      \tikzstyle{trans cost}=[draw,rectangle,minimum width=0.8em,minimum
      height=0.8em,solid,inner sep=0.07em,fill=white,font=\footnotesize];
      \tikzstyle{prob}=[inner sep=.2em, above,font=\footnotesize];

      \node[loc,accepting] (acc) at (-2.5,0) {${acc,0}$};

      \node[loc] (b0) at (0,0) {${\idle,0}$}; 
      \node[loc] (s0) at (0,-2) {${\sleep,0}$};

      \node[loc] (b1) at (2.5,0) {${\busy,1}$}; 
      \node[loc] (s1) at (2.5,-2) {${\sleep,1}$};

      \node[loc] (b2) at (5,0) {${\busy,2}$}; 
      \node[loc] (s2) at (5,-2) {${\sleep,2}$};

      \node[] (b3) at (7.5,0) {$\cdots$}; 
      \node[] (s3) at (7.5,-2) {$\cdots$};

      \node[loc] (bn) at (10,0) {$\busy, n$}; 
      \node[loc] (sn) at (10,-2) {$\sleep, n$};

\path[exp, bend left=\bend] (b1) edge node[prob, near start] {12.5}
(acc);

      \path[->,>=stealth] ($(b0)+(-0.9,0)$) edge (b0);

\path[exp, bend left=\bend] (b0) edge node[prob] {1.39} 
(b1);

\path[exp, bend left=\bend] (b1) edge node[prob] {1.39} 
(b2); 

\path[exp, bend left=\bend] (b2) edge node[prob] {12.5}
(b1);

\path[exp, bend left=\bend] (b2) edge node[prob] {1.39} 
(b3); 

\path[exp, bend left=\bend] (b3) edge node[prob] {12.5}
(b2);

\path[exp, bend left=\bend] (b3) edge node[prob] {1.39} 
(bn); 

\path[exp, bend left=\bend] (bn) edge node[prob] {12.5}
(b3);

\path[loop right,exp,looseness=5] (bn) edge node[prob, above, pos=0.2] {1.39}
(bn);

\path[exp] (s0) edge node[prob] {1.39} 
(s1); 

\path[exp] (s1) edge node[prob] {1.39} 
(s2); 

\path[exp] (s2) edge node[prob] {1.39} 
(s3); 

\path[exp] (s3) edge node[prob] {1.39} 
(sn); 

\path[loop right,exp,looseness=5] (sn) edge node[prob, above, pos=0.2] {1.39}
(sn);

\path[bend right=\bend,fixed] (b0) edge node[prob,left=-2, pos =0.25] {$1$}
(s0);

\path[bend right=\bend,fixed] (s0) edge node[prob, right, pos =0.25] {$2$}
(b0);

\path[fixed] (s1) edge node[prob,right, pos =0.25] {$2$}
(b1);

\path[fixed] (s2) edge node[prob,right, pos =0.25] {$2$}
(b2);

\path[fixed] (sn) edge node[prob,right, pos =0.25] {$2$}
(bn);

  \end{tikzpicture}
  \end{center}
  \caption{A fdCTMC model of Fujitsu disk drive}
  \label{fig:dpmsleep} 
\end{figure*}

Let us note that the ``symbolic'' algorithm could handle even larger instances, but we cannot provide such results with our current experimental implementation because of the limitation of the double precision in floating types (we would need a~higher precision). 

Table~\ref{tab:sync2} shows the timeouts synthesized for the models. As we already mentioned in Example~\ref{exa-protocol}, the timeout should depend on the number of connections that are yet to be established, so there are $n$ timeouts for a protocol involving $n$~Bobs.

\subsection{Model II., Dynamic power management of a Fujitsu disk drive}

\begin{table}[t]
\begin{center}
\begin{tabular}{| c | l  r | r | r | r | r |}
\hline
\multirow{3}{*}{ The $n$} 
& \multicolumn{6}{c|}{CPU time [s]}  \\
\cline{2-7}
& $\eps:$  & 0.005 & 0.0025 & 0.0016 & 0.00125 & 0.00100 \\
\cline{2-7} 
& $ 1/\eps:$ & 200  & 400  & 600  & 800  & 1000 \\ 
\hline
2 & \multicolumn{2}{r|}{ 17.29 } & 36.46 & 58.05 & 86.73 & 98.63 \\
4 & \multicolumn{2}{r|}{ 37.07 } & 76.60 & 133.88 & 944.08 & 1189.89 \\
6 & \multicolumn{2}{r|}{ 52.05 } & 132.18 & 1100.78 & 1336.70 & 1519.26 \\
8 & \multicolumn{2}{r|}{ 115.95 } & 1252.82 & 2321.93 & 3129.16 & 3419.42 \\
\hline
\end{tabular}

\end{center}
\caption{Running times of the ``explicit'' algorithm, Model II.}
\label{tab:pol-iter-new}
\end{table}

In this section, we consider the same simplified model of dynamic power management of a Fujitsu disk drive
that was previously analyzed\footnote{Since the implementation of the ``explicit'' algorithm was improved since the time of publishing \cite{KRF:iFM2016}, we used this new improved version in our comparisons, and hence the outcomes reported in our tables are somewhat better than the ones given in \cite{KRF:iFM2016}.} by the ``explicit'' algorithm in \cite{KRF:iFM2016}.

The model is shown in Fig.~\ref{fig:dpmsleep}. The disk has three modes $\idle$, $\busy$, and $\sleep$. In
the $\idle$ and $\sleep$ modes the disk receives requests, in the $\busy$ mode it also serves them. The disk is equipped with a bounded buffer, where it stores requests when they arrive. The requests arrive with an exponential inter-arrival time of rate $1.39$ and increase the current size of the buffer. The requests are served in an exponential time of rate $12.5$, what decreases the buffer size.  Note that restricting the model to the $\idle$ and $\busy$ modes only, we obtain a CTMC model of an M/M/1/n queue.

Moreover, the disk can move from the $\idle$ mode to the $\sleep$ mode where
it saves energy. 
Switching of the disk to the $\sleep$ mode is driven by timeout. This is
modeled by a fixed-delay transition that moves the state from $(\idle,0)$ to
$(\sleep,0)$ when the disk is steadily idle for $\timeouts((\idle,0))$
seconds. The disk is woken up by another timeout which is enabled in all
$\sleep$ states. After staying in the $\sleep$ mode for
$\timeouts((\sleep,0))$ seconds it changes the state according to the
dashed~arrows.

Note that in this example, the rates are assigned to exponential transitions, and hence the underlying CTMC
is specified by a transition matrix of rates rather than by a common exit rate $\lambda$ and a
the stochastic matrix $\prob$. Also note that the exit rates (i.e., sums of rates
on outgoing transitions) differ between $\busy$ states and the other
states. This is solved by uniformization that adds to every $\idle$ and
$\sleep$ state a self loop with rate $12.5$ and zero impulse cost.  Observe
that the introduction of exponential self loops with zero impulse cost has
no effect on the behaviour of fdCTMC including the expected cost. Now the
common exit rate $\lambda$ is $13.89$ and the stochastic matrix $\prob$ is
the transition matrix of rates multiplied by $1/\lambda$, which is the model we actually analyze.

Additionally, every state is given a rate cost that specifies an amount of
energy consumed per each second spent there.
We are interested in synthesizing optimal  timeouts for
$\timeouts((\idle,0))$ and $\timeouts((\sleep,\cdot))$ so that the average energy consumption 
before emptying the buffer is minimized.

Table~\ref{tab:pol-iter-new} and~\ref{tab:res-new} show the time needed to compute an \mbox{$\eps$-optimal} delay function for the model of Fig.~\ref{fig:dpmsleep} where $n=2,4,6,8$ and $\eps$ is progressively smaller. 
Again, the ``symbolic'' algorithm performs significantly better, especially for smaller~$\eps$ where the action space of the associated MDP $\mdp$ is already quite large.

\begin{table}[t]
\begin{center}
\begin{tabular}{| c | l  r | r | r | r | r |}
\hline
 \multirow{3}{*}{ The $n$ }
& \multicolumn{6}{c|}{CPU time [s]} \\
\cline{2-7}
& $\eps:$  & 0.005 & 0.0025 & 0.0016 & 0.00125 & 0.00100  \\
\cline{2-7}
& $ 1/\eps:$ & 200  & 400  & 600  & 800  & 1000 \\ \hline
2 & \multicolumn{2}{r|}{ 2.22 } & 2.34  & 2.34  & 2.39 & 2.42  \\
4 & \multicolumn{2}{r|}{ 2.37 } & 2.38  & 2.40  & 2.37 & 2.38  \\
6 & \multicolumn{2}{r|}{ 2.39 } & 2.39  & 2.43  & 2.39 & 2.42  \\
8 & \multicolumn{2}{r|}{ 2.40 } & 2.42  & 2.44  & 2.46 & 2.44  \\
\hline
\end{tabular}
                 
\end{center}
\caption{Running times of the ``symbolic'' algorithm, Model II.}
\label{tab:res-new}
\end{table}

\section{Conclusions}
In this paper, we designed a symbolic algorithm for fixed-delay synthesis in fdCTMC. Since the preliminary experimental results seem rather optimistic, we plan to improve our implementation and analyze the real limits of the method. To achieve that, we need  to integrate larger precision data structures and a more suitable library for root isolation.

\bibliographystyle{plain}

\end{document}